\documentclass[apj]{emulateapj}

\usepackage{natbib}
\usepackage[usenames]{color}

\newcommand{\bd}{\begin{displaymath}}
\newcommand{\ed}{\end{displaymath}}
\newcommand{\be}{\begin{equation}}
\newcommand{\ee}{\end{equation}}
\newcommand{\beaa}{\begin{eqnarray*}}
\newcommand{\eeaa}{\end{eqnarray*}}
\newcommand{\bea}{\begin{eqnarray}}
\newcommand{\eea}{\end{eqnarray}}

\def\hst{\textit{HST}}

\def\rein{r_{\rm ein}}

\def\chitah{{\sc Chitah}}
\def\glee {{\sc Glee}}

\newcommand{\sref}[1]{Section~\ref{#1}}
\newcommand{\aref}[1]{Appendix~\ref{#1}}
\newcommand{\fref}[1]{Figure~\ref{#1}}

\newcommand{\tref}[1]{Table~\ref{#1}}

\def\eg{{e.g.,}}

\begin{document}

\title{Galaxy-scale gravitational lens candidates from the Hyper Suprime-Cam imaging survey and the Galaxy And Mass Assembly spectroscopic survey}

\author{
James H.~H.~Chan\altaffilmark{1,2},
Sherry H.~Suyu\altaffilmark{3,2},
Anupreeta More\altaffilmark{4},
Masamune Oguri\altaffilmark{5,6,7},
Tzihong Chiueh\altaffilmark{1,8,9},
Jean Coupon\altaffilmark{10},
Bau-Ching~Hsieh\altaffilmark{2},
Yutaka Komiyama\altaffilmark{11,12},
Satoshi Miyazaki\altaffilmark{11,12},
Hitoshi Murayama\altaffilmark{5,11,12},
Atsushi J.~Nishizawa\altaffilmark{15},
Paul Price\altaffilmark{16},
Philip J.~Tait\altaffilmark{17},
Tsuyoshi Terai\altaffilmark{17},
Yousuke Utsumi\altaffilmark{18},
Shiang-Yu Wang\altaffilmark{2}
}
\altaffiltext{1}{Department of Physics, National Taiwan University, Taipei 10617, Taiwan}
\altaffiltext{2}{Institute of Astronomy and Astrophysics, Academia Sinica, P.O.~Box 23-141, Taipei 10617, Taiwan}
\altaffiltext{3}{Max-Planck-Institut f{\"u}r Astrophysik, Karl-Schwarzschild-Str. 1, 85741 Garching, Germany}
\altaffiltext{4}{Kavli IPMU (WPI), UTIAS, The University of Tokyo, Kashiwa, Chiba 277-8583, Japan }
\altaffiltext{5}{Kavli Institute for the Physics and Mathematics of the Universe (Kavli IPMU, WPI), University of Tokyo, 5-1-5 Kashi- wanoha, Kashiwa-shi, Chiba 277-8583, Japan}
\altaffiltext{6}{Research Center for the Early Universe, University of Tokyo, 7-3-1 Hongo, Bunkyo-ku, Tokyo 113-0033, Japan}
\altaffiltext{7}{Department of Physics, University of Tokyo, 7-3-1 Hongo, Bunkyo-ku, Tokyo 113-0033, Japan}
\altaffiltext{8}{Institute of Astrophysics, National Taiwan University, 10617 Taipei, Taiwan}
\altaffiltext{9}{Center for Theoretical Sciences, National Taiwan University, 10617 Taipei, Taiwan}
\altaffiltext{10}{Astronomical Observatory of the University of Geneva, ch. d'Ecogia 16, CH-1290 Versoix, Switzerland}
\altaffiltext{11}{University of California, Berkeley, CA 94720, USA}
\altaffiltext{12}{Lawrence Berkeley National Laboratory, MS 50A-5104, Berkeley, CA 94720. USA}
\altaffiltext{13}{National Astronomical Observatory of Japan, 2-21-1 Osawa, Mitaka, Tokyo 181-8588, Japan}
\altaffiltext{14}{Department of Astronomy, School of Science, SOKENDAI (The Graduate University for Advanced Studies), Mitaka, Tokyo 181-8588}
\altaffiltext{15}{Institute for Advanced Research, Nagoya University, Furocho Chikusa Nagoya Aichi 4648602, Japan}
\altaffiltext{16}{Princeton University Observatory, Peyton Hall, Princeton, NJ 08544, USA}
\altaffiltext{17}{National Astronomical Observatory of Japan, 650 North A'ohoku Place, Hilo, HI 96720, USA}
\altaffiltext{18}{Hiroshima Astrophysical Science Center, Hiroshima University, 1-3-1 Kagamiyama, Higashi-Hiroshima, Hiroshima 739-8526, Japan}

\email{d00222002@ntu.edu.tw}
\shorttitle{New lenses from HSC survey}
\shortauthors{Chan, Suyu, More et al.}


\begin{abstract}

We present a list of galaxy-scale lens candidates including a highly probable interacting galaxy-scale lens 
in the Hyper Suprime-Cam (HSC) imaging survey. 
We combine HSC imaging with the blended-spectra catalog from the Galaxy And Mass Assembly (GAMA) survey to identify lens candidates, 
and use lens mass modeling to confirm the candidates.
There are 45 matches between the HSC S14A\_0b imaging data release and the GAMA catalog.
We separate lens and lensed arcs using color information,
and exclude those candidates 
with small image separations ($<1.0$\arcsec, estimated with the lens/source redshifts from the GAMA survey) 
that are not easily resolved with ground-based imaging.
After excluding these, we find ten probable lens systems.
There is one system with an interacting galaxy pair, HSC J084928+000949, that has a valid mass model.
We predict the total mass enclosed by the Einstein radius of $\sim0.72$\arcsec ($\sim1.65$\,kpc) for this new expected lens system to be $\sim10^{10.59}\,M_{\odot}$.
Using the photometry in the {\it grizy} bands of the HSC survey and stellar population synthesis modeling with a Salpeter stellar initial mass function, 
we estimate the stellar mass within the Einstein radius to be $\sim10^{10.46}\,M_{\odot}$. 
We thus find a dark matter mass fraction within the Einstein radius of $\sim25\%$.
Further spectroscopy or high-resolution imaging would allow confirmation of the nature of these lens candidates.
The particular system with the interacting galaxy pair, if confirmed, would provide an opportunity to study the interplay between dark matter and stars as galaxies build up through hierarchical mergers.

\end{abstract}



\section{Introduction} 
\label{sec:intro}

Strong gravitational lensing is a powerful tool for measuring the mass distribution from galaxies to galaxy clusters.
Understanding the interplay between dark matter and baryons at galactic/sub-galactic scales is crucial to study the formation and evolution of galaxies.
Moreover, the signal from background source objects is magnified so we can make use of this information to probe the high-redshift universe.

Strong lens systems are rare, however. A lensing phenomenon happens 
when a massive foreground deflector is sufficiently well aligned along the line of sight to background source galaxies.
Searches for strongly lensed galaxies have been carried out extensively in large imaging and spectroscopic surveys, including 
the Sloan Lens ACS Survey \citep[SLACS; \eg][]{BoltonEtal06SLACS}, 
the CFHTLS\footnote{Canada-France-Hawaii Telescope Legacy Survey.  See http://www.cfht.hawaii.edu/Science/CFHLS/ and links therein for a comprehensive description.} 
Strong Lensing Legacy Survey \citep[SL2S; \eg][]{CabanacEtal07, GavazziEtal12,MoreEtal12, SonnenfeldEtal13a,SonnenfeldEtal13b},
the BOSS Emission-line Lens Survey \cite[BELLS; \eg][]{BrownsteinEtal12,BoltonEtal12a},
the \hst\ Archive Galaxy-scale Gravitational Lens Search \citep[HAGGLeS;][]{MarshallEtal09}, 
\textit{Herschel} ATLAS \citep[H-ATLAS; \eg][]{NegrelloEtal10,GonzalezNuevoEtal12}, and the South Pole Telescope \citep[SPT; \eg][]{VieiraEtal13,HezavehEtal13}. 
Through these galaxy-galaxy strong lens surveys, there are now a couple of hundreds of strong lenses with different source populations.

In this work, we start from all the
candidates in the Galaxy And Mass Assembly (GAMA) blended spectra catalog \citep{HolwerdaEtal15} 
and visually inspect their Hyper Suprime-Cam (HSC) images to prune the lens candidates further. 
With high image quality in the HSC survey, we can discern foreground lens and background source with \chitah\ \citep{ChanEtal15} using color information, 
and further reconstruct the lens mass distribution and lensed source surface brightness  
with the lens modeling software \glee\  \citep{Suyu&Halkola10, SuyuEtal12}. 
Our philosophy is to use lens modeling to confirm lens candidates (as advocated by, \eg\ \citeauthor{MarshallEtal09} \citeyear{MarshallEtal09}), 
since a lens system must be explainable by a physical lens mass model. 
Through lens mass modeling, we report ten probable lens candidates, and one of them is a highly probable interacting galaxy-scale lens system.
 
This paper is organized as follows. 
In \sref{sec:observation} we overview the observational data with spectroscopy and imaging. 
The procedure of lens examination is described in \sref{sec:method}. 
We list probable lens candidates in \sref{sec:candidates}.
In \sref{sec:summary}, conclusions are drawn and future prospect is discussed. 


\section{Observation}
\label{sec:observation}

We summarize the spectroscopic and imaging observations 
of the new lens candidates in GAMA and HSC surveys in \sref{subsec:catalog} and \sref{subsec:imaging} respectively.

\subsection{GAMA blended spectra catalog}
\label{subsec:catalog}
The GAMA survey has spectroscopic redshifts of ~300,000 galaxies down to $r<19.8$ mag over ~286 deg$^2$ 
with the upgraded 2dF spectrograph AAOmega on the Anglo-Australian Telescope 
\citep{DriverEtal09,DriverEtal11,BaldryEtal10}. 
The fiber size is 2\arcsec\ in diameter projected on the sky 
\citep{HopkinsEtal13}.
In the GAMA survey pipeline, \citet{BaldryEtal14} developed a fully automatic algorithm, AUTOZ, 
which measures redshift using a cross-correlation method for both absorption-line and emission-line features. 

An object with blended spectra is likely to be a lens candidate. 
\citet{HolwerdaEtal15} have classified the blended spectra by the best-fit passive/emission-line templates
and have listed 280 objects whose features 
of blended spectra are separated by at least 600 km/s.
They visually inspected the images of candidates in the Sloan Digital Sky Survey (SDSS) and Kilo Degreee Survey (KiDS) and obtained 
104 lens candidates and 176 occulting galaxy pairs, similar to the search in \citet{HolwerdaEtal07} in SDSS, 
supplementing eyeball searches  
in Galaxy Zoo \citep{KeelEtal13}.
The catalog provides useful information for us to further classify some of the lens candidates based on the HSC imaging survey with better image quality.

\subsection{HSC imaging}
\label{subsec:imaging}

HSC is the new prime-focus camera on the 8.2-m Subaru telescope located at Mauna Kea, Hawaii. 
This new camera together with the large telescope provides
  uniform quality of images across a wide area of the sky, i.e.,  
less affected by the vignetting and distortion around the edge of the
field of view (FOV). 
The FOV of HSC is 1.5 deg in diameter \citep{MiyazakiEtal12} which is ten times larger than the FOV of Suprime-Cam \citep{MiyazakiEtal02}. 
The HSC Subaru Strategic Program (SSP) survey consists of three layers with different depths: wide, deep and ultra-deep. 
The wide survey will cover $\sim1400$ deg$^2$ to $i\sim25.8$ mag in {\it grizy} broad bands.
We identify and model candidates in the internal early data release S14A\_0b (XMM, GAMA09h, VVDS) of the HSC imaging survey.

The data are processed with hscPipe 3.4.1, a derivative of the 
the Large Synoptic Survey Telescope (LSST) pipeline \citep{IvezicEtal08,AxelrodEtal10}, modified for use with Suprime-Cam and Hyper Suprime-Cam.
The photometric calibration is based on data obtained from the Panoramic Survey Telescope and Rapid Response System (Pan-STARRS) 1 imaging survey \citep{SchlaflyEtal12,TonryEtal12,MagnierEtal13}.
We use HSC SSP data release S15A reduced with hscPipe 3.8.5, which was released near the final stages of our analysis, 
to update our lens models of the candidates identified in S14A\_0b.
The pixel scale of the reduced images in the data releases is 0.168\arcsec.
The seeing is 0.6\arcsec in the {\it i}-band and 0.68\arcsec--0.81\arcsec in others.
\section{Method}
\label{sec:method}

\begin{figure*}
\centering
\includegraphics[scale=0.7]{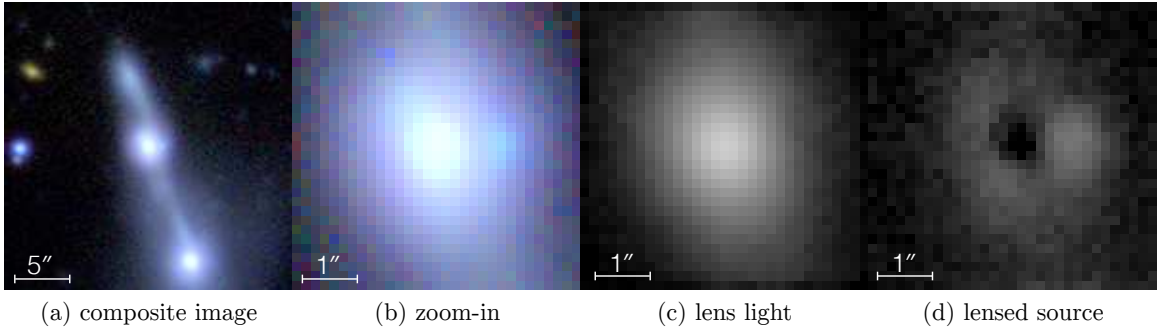}
\caption{
The highly probable lens system from HSC survey: HSC J084928+000949. 
The {\it riz} composite cutout in $25\arcsec \times 25\arcsec$ shows clearly the interacting feature in panel (a). 
Panel (b) is the central $5\arcsec \times 5\arcsec$ region of the {\it riz} composite.
\chitah\ separates the light components in panel (b) into the lens light (panel (c)) and lensed arcs of the background source (panel (d)) using color information.
The GAMA spectroscopic redshifts are $(z_1, z_2) = (0.128^{\rm P},0.603^{\rm E})$,  where P and E stand for ``Passive galaxy'' and ``Emission-line galaxies'', respectively.
}
\label{fig:new_lens}
\end{figure*}

To further confirm the lens candidates in the GAMA blended spectra catalog, we visually select objects based on their morphology and redshift information.
Moreover, we try to reconstruct background sources via lens modeling.
If a candidate is a lens system, its reconstructed source components should be well described by regular looking galaxies.
In other words, we can confirm a system to be a lens if we can find a physically sensible model 
for the foreground lens mass distribution and the background source surface brightness.

\subsection{Selection via morphology and redshift information}
\label{subsec:selection}

We start with the 280 objects from the GAMA blended spectra catalog. 
There are 45 objects in the footprint of the HSC early data release S14A\_0b.
We exclude two of them with incomplete HSC imaging data.
In the remaining 43 objects, there are 14 objects identified as strong
lens candidates in \citet{HolwerdaEtal15}. 
We note that these objects are categorized by \citet{HolwerdaEtal15}
as having a passive galaxy (PG) at a lower redshift near an
emission-line galaxy (ELG) at a higher redshift.  In addition to these
types of ``PG+ELG'' blended spectra (foreground passive galaxy with
background emission-line galaxy), there are the ``PG+PG'' (foreground
and background passive galaxies), ``ELG+PG'' (foreground emission-line
galaxy with background passive galaxy) and ``E+E'' (foreground and
background emission-line galaxies) types identified by
\citet{HolwerdaEtal15} that could also be potential strong lenses.
Therefore, we consider all the 43 galaxies from \citet{HolwerdaEtal15}
with HSC imaging in the rest of the paper.  

We employ \chitah\ \citep{ChanEtal15} to disentangle lens and lensed-source components based on color information.
\chitah\ combines two image cutouts, one from the bluer bands ({\it g} or {\it r}) and one from the redder bands ({\it z} or {\it y}), to produce a ``foreground lens image'' and a ``lensed arc image of the background source''. 
\chitah\ chooses the band with a narrower PSF from the two bluer bands, and similarly from the two redder bands.
By selecting the cutouts with narrower PSFs, \chitah\ produces the sharpest possible images of the lens and lensed arcs.
We show in \fref{fig:new_lens}(c) and (d) an example of the lens light and lensed arc separation, respectively.

Furthermore, \citet{HolwerdaEtal15} also provided two galaxies' redshifts of each lens candidate 
which allow us to predict image separations, $\vartheta_{\rm sep}$, with a fiducial velocity dispersion of $\sigma\sim250$\,km/s, 
\be
  \vartheta_{\rm sep} = 4 \pi {\sigma^2 D_{\rm LS} \over c^2 D_{\rm S}},
\ee
where $D_{\rm LS}$ is the angular diameter distance between the lens and the source, and $D_{\rm S}$ is the angular distance between us and the source. 
Those lens candidates with lens/source redshift values that are too close are excluded 
since the image separations are too small ($\vartheta_{\rm sep} < 1.0\arcsec$) and difficult to resolve in the HSC images. 
We also empirically rule out the candidates that have no lensed arc features 
and only 10 candidates remain. 
In other words, when the lensed-arc feature can be seen such as that shown in \fref{fig:new_lens}(d), 
we then try to model the lens mass distribution and source surface brightness of the system.

\subsection{Lens model}
\label{subsec:model}

To confirm the lensing nature of the 10 probable candidates, we use the lens modeling software \glee, developed by A.~Halkola and S.~H.~Suyu \citep{Suyu&Halkola10, SuyuEtal12},
to fit the lens light and lensed-source components. 
Lens light components show stellar distributions of lens galaxies, which can be well described by S{\'e}rsic profiles \citep{Sersic63}:
\be
\label{eq:sersic}
I(\theta_1,\theta_2) = A \exp \left[ -k \left (\left(\frac{\sqrt{\theta_1^2 +
      \theta_2^2/q_{\rm L}^2}}{R_{\rm eff}}\right) ^ {1/n_{\rm
    s}} - 1 \right) \right],
\ee
where $(\theta_1,\theta_2)$ are image coordinates, $A$ is the amplitude, $k$ is a constant such that $R_{\rm eff}$ is the effective (half-light) radius, $q_{\rm L}$ is the axis ratio, and $n_{\rm s}$ is the S{\'e}rsic index.  To fit to the lens light image from \chitah, we have as additional parameters the centroid $(\theta_{\rm 1c},\theta_{\rm 2c})$ and the position angle of the S{\'e}rsic distribution.
For the lens galaxy's dimensionless surface mass density (the total mass distribution, including baryons and dark matter), 
we adopt pseudo-isothermal elliptic mass distribution \citep[PIEMD;][]{Kassiola&Kovner93} with a vanishing core:
\be
\label{eq:piemd}
\kappa(\theta_1, \theta_2) = \frac{\rein}{2\sqrt{\frac{{\theta_1^2}}{(1+\epsilon)^2}+\frac{\theta_2^2}{(1-\epsilon)^2}}},
\ee
where $\rein$ is the Einstein radius of the lens (that is related to the strength of the lens) and $\epsilon$ is the ellipticity that is related to the axis ratio $q$ by $\epsilon \equiv (1-q)/(1+q)$.  
 
We first model the lens light produced in the color-difference image as shown in \fref{fig:new_lens}(c) to obtain the S{\'e}rsic parameters, particularly the centroid and position angle. 
We then model the lensed arcs to constrain simultaneously the source surface brightness on a grid of pixels and the lens mass parameters of PIEMD.  
We adopt the curvature form of regularization on the source grid, 
and Gaussian priors on the centroid ($\sim$0.05\arcsec), the axis ratio ($\sim$0.1), and the  position angle ($\sim$3 deg) of the PIEMD based on the fitted S{\'e}rsic parameters.  
We allow the Einstein radius to vary with uniform prior.
We use \glee\ to confirm the lensing nature for those systems that have sensible source surface brightness and lens mass model parameters, as detailed next.

\section{Gravitational lens candidates from HSC survey}
\label{sec:candidates}
Of the ten candidates that show arc-like features in
  \sref{subsec:selection}, we 
cannot obtain reasonable lens models with \glee\ for nine of them, due to e.g.~$\rein$ that are too small/large or inverted sources that appear unphysical as a result of the limited number of image pixels.
We list these nine lens candidates in \sref{subsec:all_cand}.
We consider one candidate that is the most probable since we can obtain a sensible lens model with \glee\ and discuss it in \sref{subsec:newlens}.
The remaining 33 objects classified as possible lenses or non-lenses are
  listed in \aref{app:table_rest}.

\subsection{Probable lens candidates}   
\label{subsec:all_cand}

\begin{figure*}
\centering
\includegraphics[scale=1.0]{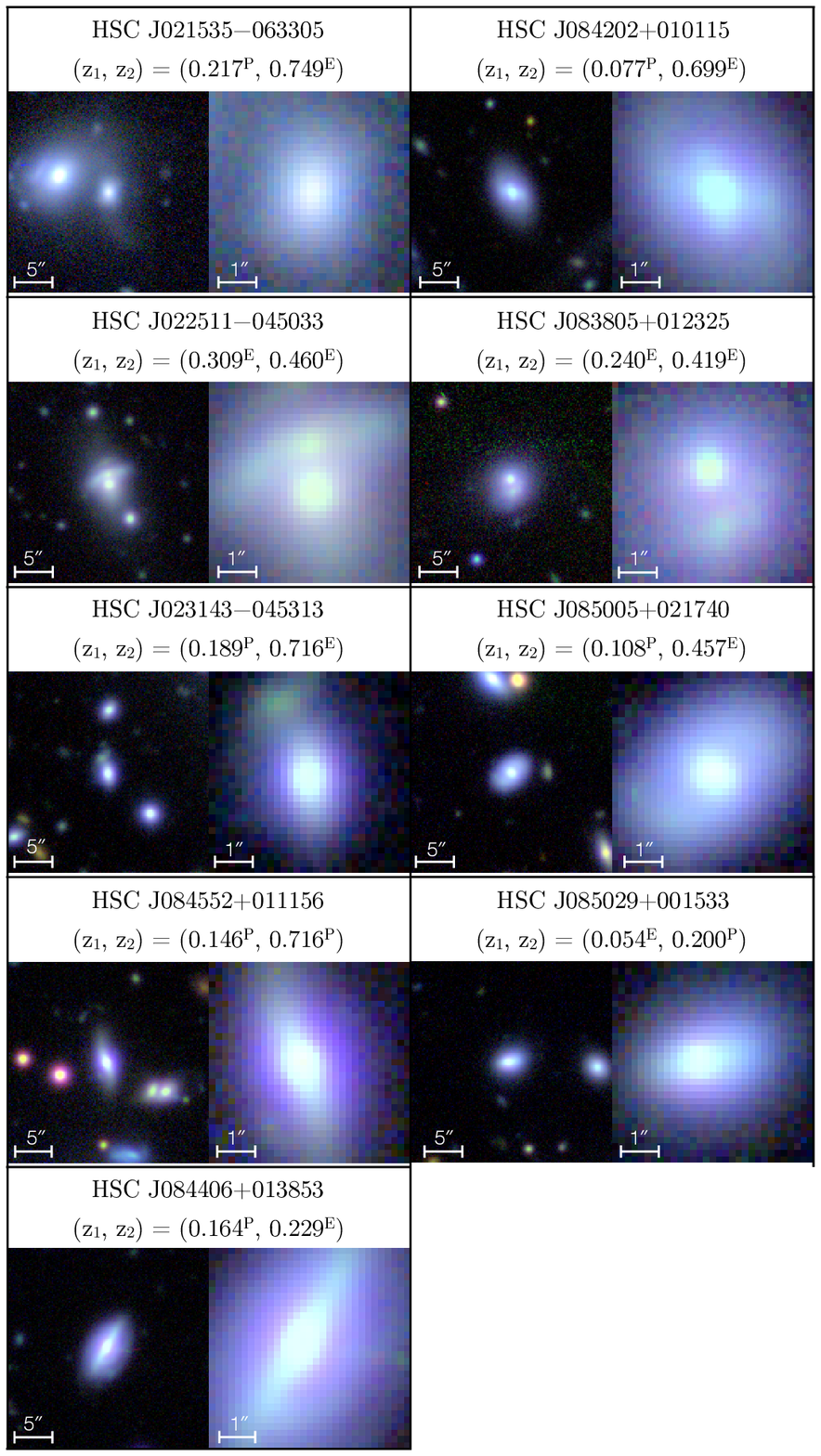}
\caption{The nine probable lens candidates from the combination of GAMA spectroscopy and HSC SSP imaging.
For each candidate, we show on the left a 25\arcsec$\times$25\arcsec cutout and on the right a 5\arcsec$\times$5\arcsec cutout of the {\it riz} color image.
The spectral types of each candidate is denoted after the redshifts, where ``P'' stands for ``passive galaxies'' and ``E'' stands for ``emission-line galaxies''. 
J022511$-$045033 is a probable binary lens system. 
J084552$+$011156 and J084406$+$013853 are likely disky lenses.
Some of them are low-redshift lenses, such as J084202$+$010115, J085005$+$021740 and J085029$+$001533. 
}
\label{fig:candidates}
\end{figure*}

The lens candidates are listed in \fref{fig:candidates} and \tref{tab:list} for future confirmation.
Except for HSC J084928+000949 which we discuss in more detail in \sref{subsec:newlens}, we classify the remaining nine systems as probable lens candidates. 
Five of these nine probable lens candidates were previously identified by
  \citet{HolwerdaEtal15} as possible strong lenses with foreground
  passive lens galaxies and background emission-line source galaxies
  (i.e., the six objects in \fref{fig:candidates} with ``P'' and ``E''
  flagged on $z_1$ and $z_2$, respectively). 

We identify a probable binary lens system in J022511$-$045033. 
J084552$+$011156 and J084406$+$013853 are likely disky lenses.
Some of the candidates are low-redshift lenses, such as J084202$+$010115, J085005$+$021740 and J085029$+$001533.
These candidates are worth inspecting with higher-resolution 
imaging, as they complement nicely the existing samples of lenses at higher redshifts, \eg\ SLACS, SL2S and BELLS.

\begin{table*}
\centering
\caption{The catalog of lens candidates selected from GAMA in the HSC survey} 
\label{tab:list}
\begin{center}
\begin{tabular}{lrrrc}
\hline
 HSC ID & GAMA ID & RA & DEC & ($z_1,z_2$) \\ \hline
HSC J084928$+$000949\textsuperscript{\S}\textsuperscript{\dag} 
                                         &  209222  &$132.36771$ & $ 0.16360$&  ($0.128^{\rm P}$, $0.603^{\rm E}$)\\
HSC J021535$-$063305\textsuperscript{\S} &  1126606 &$ 33.89513$ & $-6.55139$&  ($0.217^{\rm P}$, $0.749^{\rm E}$)\\
HSC J022511$-$045033                     &  1779869 &$ 36.29463$ & $-4.84248$&  ($0.309^{\rm E}$, $0.460^{\rm E}$)\\
HSC J023143$-$045313\textsuperscript{\S} &  2005629 &$ 37.92875$ & $-4.88681$&  ($0.189^{\rm P}$, $0.716^{\rm E}$)\\
HSC J084552$+$011156                     &  300979  &$131.46746$ & $ 1.19884$&  ($0.146^{\rm P}$, $0.716^{\rm P}$)\\
HSC J084406$+$013853\textsuperscript{\S} &  323247  &$131.02333$ & $ 1.64814$&  ($0.164^{\rm P}$, $0.229^{\rm E}$)\\
HSC J084202$+$010115\textsuperscript{\S} &  371208  &$130.50925$ & $ 1.02071$&  ($0.077^{\rm P}$, $0.699^{\rm E}$)\\
HSC J083805$+$012325                     &  375506  &$129.51933$ & $ 1.39028$&  ($0.240^{\rm E}$, $0.419^{\rm E}$)\\
HSC J085005$+$021740\textsuperscript{\S} &  417645  &$132.51888$ & $ 2.29438$&  ($0.108^{\rm P}$, $0.457^{\rm E}$)\\
HSC J085029$+$001533                     &  599797  &$132.62096$ & $ 0.25907$&  ($0.054^{\rm E}$, $0.200^{\rm P}$)\\ 
\hline                           
\end{tabular}
\end{center}
\S The strong lens candidate in GAMA blended spectra catalog.
\dag The highly probable lens system via lens modeling.
The Einstein radius is $\rein\sim0.72\arcsec$. 
The total mass within the Einstein radius is $M_{\rm tot} \sim 10^{10.59} M_{\odot}$. 
The stellar mass is $M_* \sim 10^{10.46} M_{\odot}$, 
accounting for dust extinction.
{\bf The remaining nine lens candidates without \dag\ cannot obtain reasonable lens models with GLEE, \eg\ too small/large $\rein$ or unphysical inverted source, 
due to the limited number of image pixels.}
{\bf The spectral types of each candidate is denoted after the redshifts, where ``P'' stands for ``passive galaxies'' and ``E'' stands for ``emission-line galaxies''.}
\end{table*}

\subsection{A promising interacting-galaxy lens candidate from HSC survey}
\label{subsec:newlens}

\begin{figure*}
\centering
\includegraphics[angle=-90,scale=0.7]{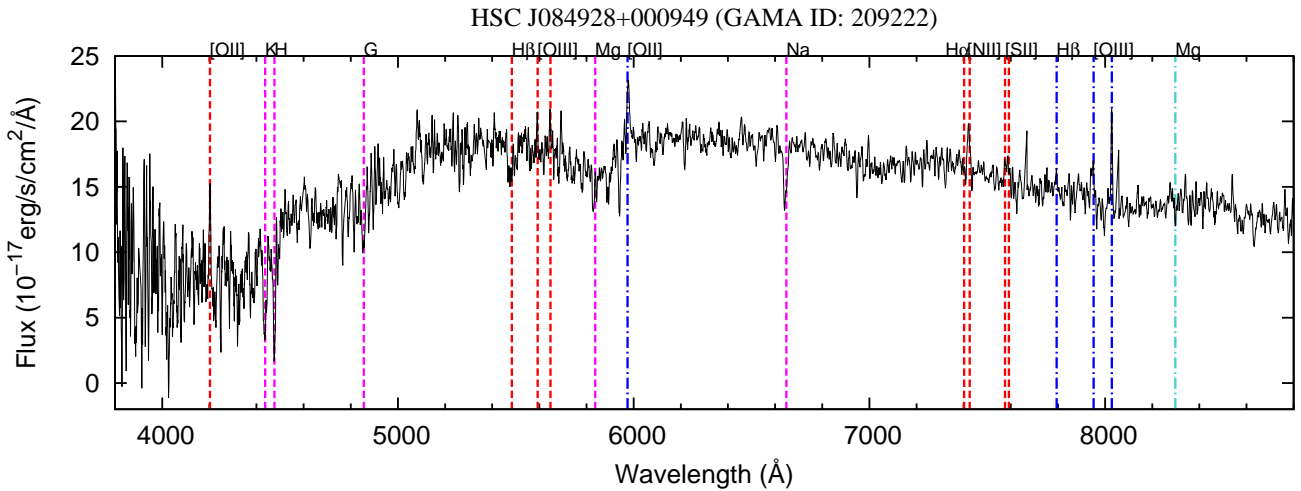}
\caption{
Blended spectra of HSC J084928+000949 (black line) from the GAMA survey (GAMA ID: 209222). Red and magenta dashed lines show the emission and absorption lines at $z_1 = 0.128$ respectively. 
Blue and cyan dot-dashed lines show the emission and absorption lines at $z_2 = 0.603$ respectively.
{\bf http://www.gama\=/survey.org/dr2/data/spectra/gama/reduced\_08/1d/G09\_Y1\_CS2\_288.fit}
}
\label{fig:spec}
\end{figure*}

In this work, we identify one very promising lens candidate, HSC J084928+000949, from HSC survey via lens modeling.
The result is shown in \fref{fig:new_lens}.
\chitah\ separates lens light and lensed arcs using {\it g}-band cutout instead of {\it r}-band cutout which has bad pixels,
even though the {\it r}-band PSF is narrower than that of the {\it g}-band.
The coordinate is (RA, DEC) = (132.36771, 0.16360). 
We also notice that this candidate is an interacting galaxy-pair which is an important aspect of galaxy evolution \citep[\eg][]{HopkinsEtal06,LotzEtal08}.
Gravitational lensing provides an independent constraint on dark matter distribution. 
Hence, it can help us to study the interplay between dark matter and baryonic components \citep[\eg][]{BarnabeEtal11,SonnenfeldEtal15}. 

We also check the GAMA spectra shown in \fref{fig:spec} \citep{LiskeEtal15} to verify that it is blended of two galaxies at different redshifts.
The two galaxy types are emission line galaxy ($z_2\sim0.603$) and passive galaxy ($z_1\sim0.128$) -- in fact, this system was previously identified by \citet{HolwerdaEtal15} as a lens candidate containing a foreground passive galaxy with a background emission-line galaxy. 
We indicate the emission and absorption lines at $z_1 = 0.128$ as red and magenta dashed lines, respectively, 
and the emission and absorption lines at $z_2 = 0.603$ as blue and cyan dot-dashed lines, respectively.  
We also note that this probable lens galaxy appears to be merging with another one in \fref{fig:new_lens}(a).

\begin{figure}
\centering
\includegraphics[scale=0.4]{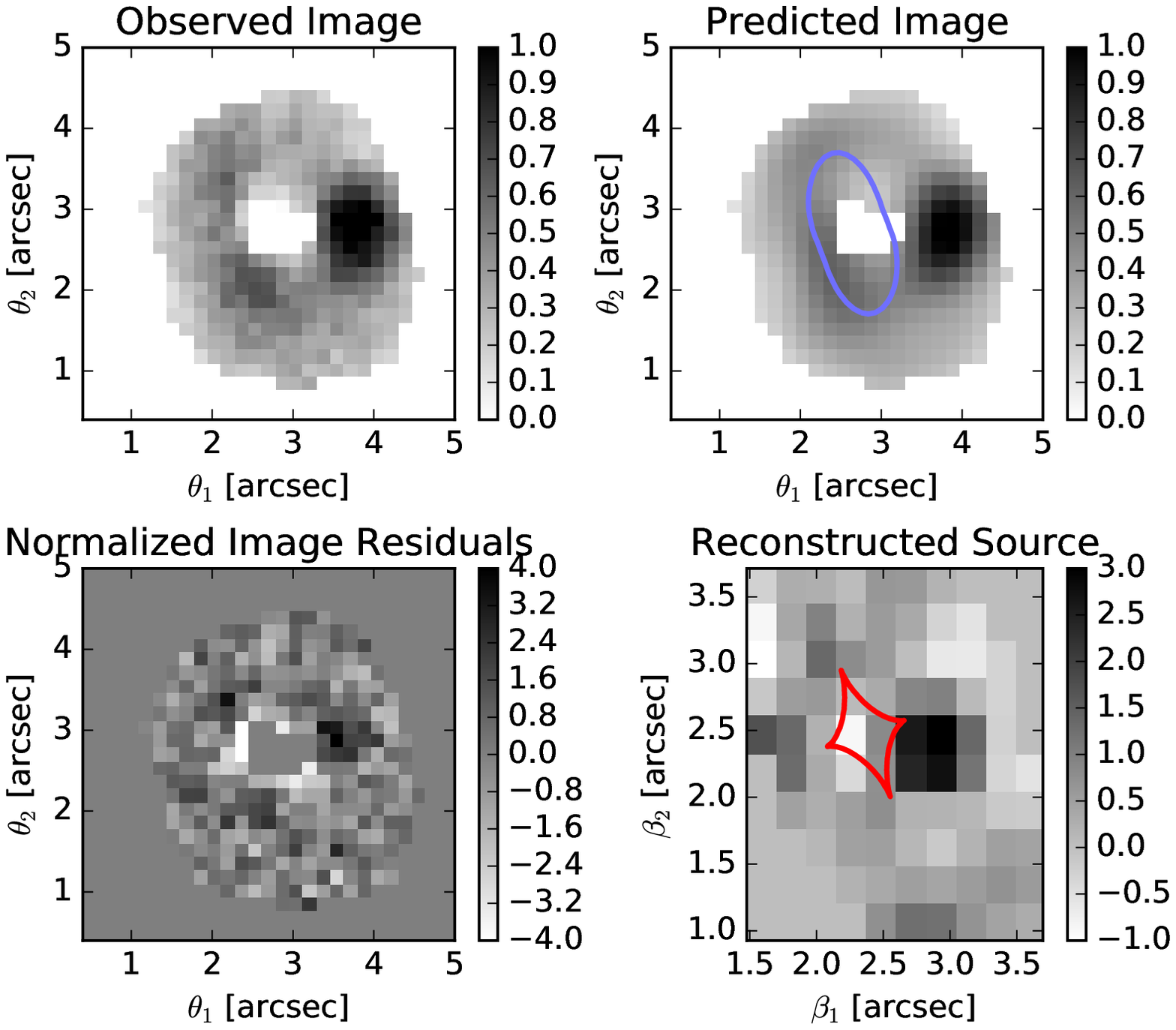}
\caption{The result from lens modeling of the highly probable lens HSC J084928+000949. We model the lensed-arc feature in the masked region 
shown in the top-left panel. 
The predicted lensed-arc of the best-fitting model is shown in the top-right panel. The normalized image residuals are shown in the bottom-left panel.
We can see clearly a compact source galaxy in the bottom-right panel. 
The critical line (blue) and the caustics (red) predicted by our lens model are drawn in the top-right and bottom-right panels, respectively. 
} 
\label{fig:lens_model}
\end{figure}

The result from \glee\ is shown in \fref{fig:lens_model}. 
We model the lensed-arc feature in the masked region shown in the top-left panel. 
The predicted lensed arc of the best fitting model is shown in the top-right panel and the normalized image residuals are shown in the bottom-left panel.
We can clearly see a single source in the bottom-right panel based on the lens model with 
$\rein = 0\arcsec.72\pm^{0.04}_{0.02}$.
The critical line (blue) and the caustics (red) predicted by our lens model are drawn in the top-right and bottom-right panels, respectively. 
Even though we have a sensible lens mass model of J084928+000949, we classify this object as a highly probably lens system rather than a definite lens. 
Spectroscopically there are two 
distinct redshifts identified. 
However, since the GAMA spectroscopy does not have sufficient spatial resolution and the foreground lens galaxy also shows emission lines,  
it is not fully clear whether the background source has multiply lensed images. 
For the purpose of lens modeling, we have decided to use certain bluish features as lensed images of the background source, 
but high-resolution imaging or spatially resolved spectroscopy is needed to unambiguously confirm this potential lens system.

Based on our current best-fit model of J084928$+$000949, we compute the total mass within the Einstein radius through
$M_{\rm tot}(<\rein)= \pi (D_{\rm L}\rein)^2 \Sigma_{\rm cr}$,
where $D_{\rm L}$ is the angular distance between us and the lens, and $\Sigma_{\rm cr} = c^2 D_{\rm L}/4\pi G D_{\rm LS}D_{\rm S}$ is the critical surface mass density. 
The total mass for this highly probable lens system is $\log(M_{\rm tot}(<\rein)/M_\odot)=10.59\pm^{0.05}_{0.02}$.
We estimate the AB magnitudes 
in all {\it grizy} bands of the lens galaxy within a circular aperture of radius $\rein$:  
$m_g=20.66\pm0.02$, $m_r=19.53\pm0.02$, $m_i=18.98\pm0.02$, $m_z=18.57\pm0.03$, and $m_y=18.47\pm0.02$.  
With the same method in \cite{OguriEtal14} using \cite{Bruzual&Charlot03} stellar population synthesis model,
we calculate the stellar mass 
$\log(M_*/M_\odot)=10.46\pm0.08$
including additional parameter on dust extinction at the galaxy
redshift. The value is derived using Salpeter initial mass funcion
(IMF). We note that we obtain similar stellar mass of
$\log(M_*/M_\odot)=10.50\pm0.08$
even if we do not include the dust extinction.
The stellar mass is consistent with the result from \citet{TaylorEtal11}, $\log(M_*/M_\odot)=10.55\pm0.11$. 
From this analysis we can infer the dark matter fraction within the Einstein radius to be $\sim 25\%$, which is comparable to those
in SLACS and SL2S \citep{BarnabeEtal11,AugerEtal10,SonnenfeldEtal15}.

\section{Summary and Discussions}
\label{sec:summary}

Starting from the GAMA blended spectra catalog, we find 45 matches in HSC early data release S14A\_0b.
We use \chitah\ to separate lens light and lensed arc components. 
We identified 10 lens candidates via redshift information and morphology.
To further confirm these lens candidates, we use \glee\ to model the lensed arc components.
After our examinations of the modeling results, we draw the following conclusions:

\begin{enumerate}
  \item  We identify 10 lens candidates via redshift information and morphology. 
  \item  Through lens model fitting and examining the spectrum, we consider one system, HSC J084928+000949, as a highly probable lens. The Einstein radius $\rein$ is around 0.72\arcsec.
  \item  The highly probable lens is an interacting system which allows us to study the interplay between dark matter and stars.
  \item  For the highly probable lens, the dark matter fraction within $\rein$ is $\sim 25\%$.
\end{enumerate}

Of the 14 strong-lens candidates from \citet{HolwerdaEtal15} that have HSC imaging in the S14A\_0b data release, we have classified 
6 of them as probable lenses.
 The success rate of the GAMA's approach is thus roughly $\sim 50\%$, 
although we note that further confirmation of these candidates is needed in order to obtain more robust estimates of the success rate. 

After the first successful identification of a list of lens candidates including one highly probable lens system, we expect to 
find  more strongly lensed galaxies in future data releases of HSC. 
Higher-resolution imaging (via, \eg\ adaptive optics), or spatially resolved spectroscopy would be helpful to confirm further the nature of these candidates.
The new interacting lens candidate is very valuable and with follow-up observation it would provide a great opportunity to study 
the interplay between dark matter and also the hierarchical formation/evolution of galaxies.


\section*{Acknowledgments}

We thank Ying-Tung Chen and Li-Hwai Lin for useful discussions,
and the anonymous referee for helpful comments.
J.H.H.C.~would like to thank Chih-Fan Chen and Kenneth Wong for algorithm support.
J.H.H.C.~and S.H.S.~gratefully acknowledge support by the Ministry of Science and Technology in Taiwan via grant MOST-103-2112-M-001-003-MY3, and support by the Max Planck Society through the Max Planck Research Group for S.H.S.
T.C.~acknowledges the Ministry of Science and Technology in Taiwan via grant MOST-103-2112-M-002-020-MY3.
A.M.~is supported by World Premier International Research Center Initiative (WPIInitiative), MEXT, Japan, 
and also acknowledges the support of the Japan Society for Promotion of Science (JSPS) fellowship.
M.O.~acknowledges support in part by World Premier International
Research Center Initiative (WPI Initiative), MEXT, Japan, and
Grant-in-Aid for Scientific Research from the JSPS (26800093).
The Hyper Suprime-Cam (HSC) collaboration includes the astronomical
communities of Japan and Taiwan, and Princeton University. The HSC
instrumentation and software were developed by the National
Astronomical Observatory of Japan (NAOJ), the Kavli Institute for the
Physics and Mathematics of the Universe (Kavli IPMU), the University
of Tokyo, the High Energy Accelerator Research Organization (KEK), the
Academia Sinica Institute for Astronomy and Astrophysics in Taiwan
(ASIAA), and Princeton University. Funding was contributed by the FIRST 
program from Japanese Cabinet Office, the Ministry of Education, Culture, 
Sports, Science and Technology (MEXT), the Japan Society for the 
Promotion of Science (JSPS), Japan Science and Technology Agency 
(JST), the Toray Science Foundation, NAOJ, Kavli IPMU, KEK, ASIAA,  
and Princeton University.
This paper makes use of software developed for the Large Synoptic Survey Telescope. We thank the LSST Project for making their code available as free software at http://dm.lsstcorp.org.
The Pan-STARRS1 Surveys (PS1) have been made possible through contributions of the Institute for Astronomy, the University of Hawaii, the Pan-STARRS Project Office, the Max-Planck Society and its participating institutes, the Max Planck Institute for Astronomy, Heidelberg and the Max Planck Institute for Extraterrestrial Physics, Garching, The Johns Hopkins University, Durham University, the University of Edinburgh, Queen's University Belfast, the Harvard-Smithsonian Center for Astrophysics, the Las Cumbres Observatory Global Telescope Network Incorporated, the National Central University of Taiwan, the Space Telescope Science Institute, the National Aeronautics and Space Administration under Grant No. NNX08AR22G issued through the Planetary Science Division of the NASA Science Mission Directorate, the National Science Foundation under Grant No. AST-1238877, the University of Maryland, and Eotvos Lorand University (ELTE).
%


\appendix
\section{The possible lens candidates from GAMA in the HSC survey}
\label{app:table_rest}

\begin{figure*}
\centering
\includegraphics[scale=1.2]{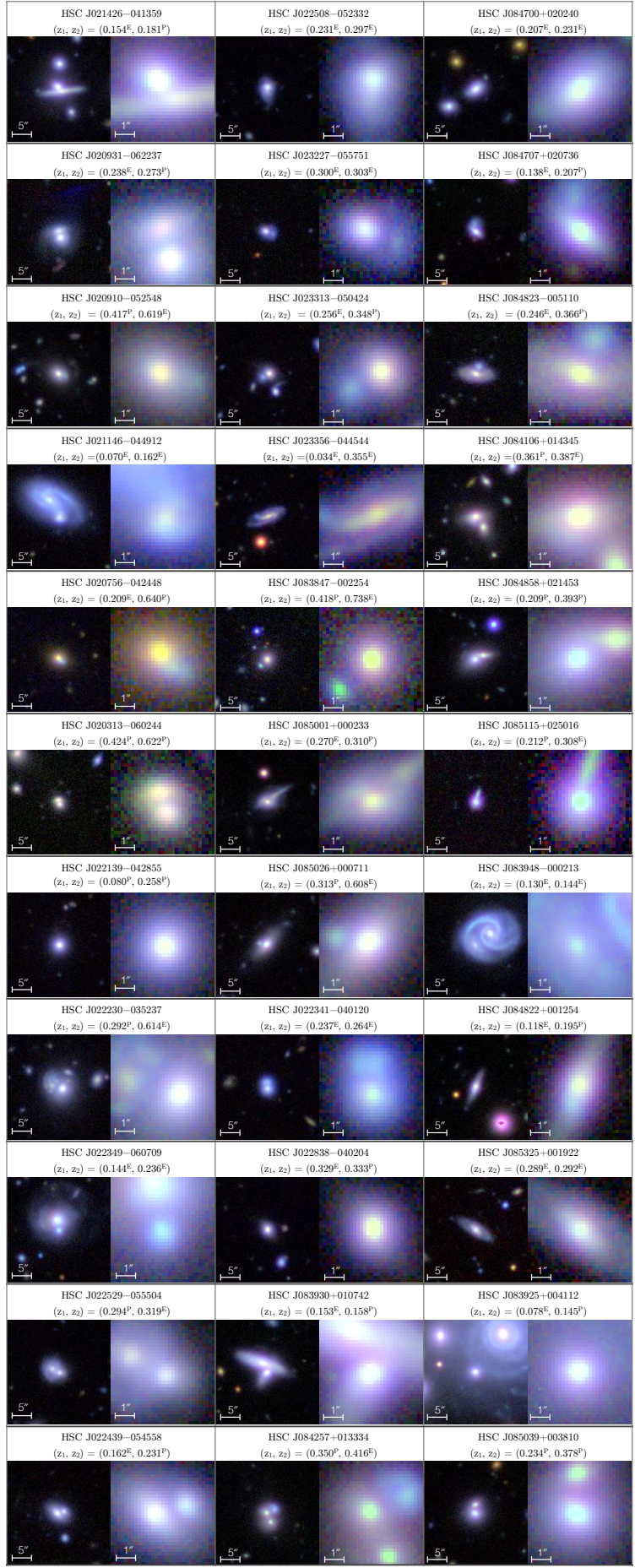}
\caption{The remaining candidates from the combination of GAMA spectroscopy and HSC SSP imaging.
For each candidate, we show on the left a 25\arcsec$\times$25\arcsec cutout and on the right a 5\arcsec$\times$5\arcsec cutout of the {\it riz} color image.
The spectral types of each candidate is denoted after the redshifts, where ``P'' stands for ``passive galaxies'' and ``E'' stands for ``emission-line galaxies''.
}
\label{fig:rest}
\end{figure*}

We classify the remaining 33 objects as possible lens candidates or non-lenses (\eg\ galaxy pairs), and list them in \tref{tab:list_remain} and \fref{fig:rest}.
Possible reasons are included in \tref{tab:list_remain}.
We mark 8 objects by \S\ that are previously identified as strong lens candidates by \citet{HolwerdaEtal15}, i.e., objects with blended GAMA spectra of foreground passive galaxies with background emission-line galaxy.
For those object with $\vartheta_{\rm sep} < 1.0$, 
higher resolution imaging from space or adaptive optics are needed
to detect lensed features.

\begin{table*}
\centering
\caption{The catalog of the possible lens candidates or non-lenses from GAMA in the HSC survey} 
\label{tab:list_remain}
\begin{center}
\begin{tabular}{lrrrcl}
\hline
 HSC ID & GAMA ID & RA & DEC & ($z_1,z_2$) &  possible reason \\\hline
HSC J021426$-$041359                     & 1217811&$  33.60950$&$ -4.23305$&($0.154^{\rm E}$, $0.181^{\rm P}$)& $\vartheta_{\rm sep} < 1.0\arcsec$     \\ 
HSC J020931$-$062237                     & 1298084&$  32.37917$&$ -6.37684$&($0.238^{\rm E}$, $0.273^{\rm P}$)& $\vartheta_{\rm sep} < 1.0\arcsec$     \\ 
HSC J020910$-$052548\textsuperscript{\S} & 1312058&$  32.29279$&$ -5.43011$&($0.417^{\rm P}$, $0.619^{\rm E}$)& no visible strong lensing feature\ddag \\ 
HSC J021146$-$044912                     & 1320592&$  32.94087$&$ -4.82006$&($0.070^{\rm E}$, $0.162^{\rm E}$)& galaxy pair              \\ 
HSC J020756$-$042448                     & 1440776&$  31.98304$&$ -4.41341$&($0.209^{\rm E}$, $0.640^{\rm P}$)& galaxy pair              \\ 
HSC J020313$-$060244                     & 1537351&$  30.80412$&$ -6.04563$&($0.424^{\rm P}$, $0.622^{\rm P}$)& galaxy pair              \\ 
HSC J022139$-$042855                     & 1675035&$  35.41175$&$ -4.48197$&($0.080^{\rm E}$, $0.258^{\rm P}$)& no resolvable strong lensing feature\ddag  \\ 
HSC J022230$-$035237\textsuperscript{\S} & 1684064&$  35.62492$&$ -3.87688$&($0.292^{\rm P}$, $0.614^{\rm E}$)& multiple objects       \\ 
HSC J022349$-$060709                     & 1760310&$  35.95496$&$ -6.11922$&($0.144^{\rm E}$, $0.236^{\rm E}$)& galaxy pair              \\ 
HSC J022529$-$055504\textsuperscript{\S} & 1763319&$  36.37075$&$ -5.91766$&($0.294^{\rm P}$, $0.319^{\rm E}$)& $\vartheta_{\rm sep} < 1.0\arcsec$     \\ 
HSC J022439$-$054558                     & 1765570&$  36.16213$&$ -5.76612$&($0.162^{\rm E}$, $0.231^{\rm P}$)& galaxy pair              \\ 
HSC J022508$-$052332                     & 1771132&$  36.28483$&$ -5.39210$&($0.231^{\rm E}$, $0.297^{\rm E}$)& $\vartheta_{\rm sep} < 1.0\arcsec$     \\ 
HSC J023227$-$055751                     & 1988308&$  38.11142$&$ -5.96424$&($0.300^{\rm E}$, $0.303^{\rm E}$)& $\vartheta_{\rm sep} < 1.0\arcsec$     \\ 
HSC J023313$-$050424                     & 2002618&$  38.30588$&$ -5.07329$&($0.256^{\rm E}$, $0.348^{\rm P}$)& $\vartheta_{\rm sep} < 1.0\arcsec$     \\ 
HSC J023356$-$044544                     & 2007752&$  38.48246$&$ -4.76219$&($0.034^{\rm E}$, $0.355^{\rm E}$)& no counter image\ddag            \\ 
HSC J083847$-$002254\textsuperscript{\S} & 202448 &$ 129.69546$&$ -0.38179$&($0.418^{\rm P}$, $0.738^{\rm E}$)& galaxy pair              \\ 
HSC J085001$+$000233                     & 209263 &$ 132.50596$&$  0.04250$&($0.270^{\rm E}$, $0.310^{\rm P}$)& $\vartheta_{\rm sep} < 1.0\arcsec$     \\ 
HSC J085026$+$000711\textsuperscript{\S} & 209295 &$ 132.61013$&$  0.11972$&($0.313^{\rm P}$, $0.608^{\rm E}$)& galaxy pair              \\ 
HSC J022341$-$040120                     & 2308869&$  35.92083$&$ -4.02211$&($0.237^{\rm E}$, $0.264^{\rm E}$)& $\vartheta_{\rm sep} < 1.0\arcsec$     \\ 
HSC J022838$-$040204                     & 2379807&$  37.16025$&$ -4.03447$&($0.329^{\rm E}$, $0.333^{\rm P}$)& $\vartheta_{\rm sep} < 1.0\arcsec$     \\ 
HSC J083930$+$010742                     & 300500 &$ 129.87425$&$  1.12844$&($0.153^{\rm E}$, $0.158^{\rm P}$)& $\vartheta_{\rm sep} < 1.0\arcsec$     \\ 
HSC J084257$+$013334\textsuperscript{\S} & 323200 &$ 130.73717$&$  1.55957$&($0.350^{\rm P}$, $0.416^{\rm E}$)& $\vartheta_{\rm sep} < 1.0\arcsec$     \\ 
HSC J084700$+$020240                     & 345974 &$ 131.75004$&$  2.04437$&($0.207^{\rm E}$, $0.231^{\rm E}$)& $\vartheta_{\rm sep} < 1.0\arcsec$     \\ 
HSC J084707$+$020736                     & 345984 &$ 131.78108$&$  2.12672$&($0.138^{\rm E}$, $0.207^{\rm P}$)& galaxy pair              \\ 
HSC J084823$-$005110                     & 3624490&$ 132.09767$&$ -0.85274$&($0.246^{\rm E}$, $0.366^{\rm P}$)& galaxy pair              \\ 
HSC J084106$+$014345\textsuperscript{\S} & 380839 &$ 130.27629$&$  1.72911$&($0.361^{\rm P}$, $0.387^{\rm E}$)& $\vartheta_{\rm sep} < 1.0\arcsec$     \\ 
HSC J084858$+$021453                     & 386427 &$ 132.24292$&$  2.24817$&($0.209^{\rm P}$, $0.393^{\rm P}$)& galaxy pair              \\ 
HSC J085115$+$025016\textsuperscript{\S} & 422882 &$ 132.81062$&$  2.83784$&($0.212^{\rm P}$, $0.308^{\rm E}$)& galaxy pair              \\ 
HSC J083948$-$000213                     & 573744 &$ 129.95100$&$ -0.03694$&($0.130^{\rm E}$, $0.144^{\rm E}$)& $\vartheta_{\rm sep} < 1.0\arcsec$     \\ 
HSC J084822$+$001254                     & 599598 &$ 132.08975$&$  0.21503$&($0.118^{\rm E}$, $0.195^{\rm P}$)& no counter image\ddag      \\ 
HSC J085325$+$001922                     & 599995 &$ 133.35229$&$  0.32283$&($0.289^{\rm E}$, $0.292^{\rm E}$)& $\vartheta_{\rm sep} < 1.0\arcsec$     \\ 
HSC J083925$+$004112                     & 621991 &$ 129.85479$&$  0.68659$&($0.078^{\rm E}$, $0.145^{\rm P}$)& no visible strong lensing feature\ddag   \\ 
HSC J085039$+$003810                     & 622326 &$ 132.66296$&$  0.63611$&($0.234^{\rm P}$, $0.378^{\rm P}$)& galaxy pair              \\ 
\hline
\end{tabular}
\end{center}
\S Strong-lens candidates in GAMA blended spectra catalog.
\ddag We inspect for strong lensing features using the ``lensed images" based on color separation by \chitah.
We predict $\vartheta_{\rm sep}$ with a fiducial velocity dispersion of $\sigma\sim250$km/s.
{\bf The spectral types of each candidate is denoted after the redshifts, where ``P'' stands for ``passive galaxies'' and ``E'' stands for ``emission-line galaxies''.}
\end{table*}

\bibliography{GamaPaper.bib}
\bibliographystyle{apj}


\label{lastpage}
\end{document}